\documentclass[showpacs,aps, prl, twocolumn]{revtex4}
\usepackage{graphics}
\usepackage{epsfig}

\begin{document}

\title{Direct observation of 4$^+$ to 2$^+$ gamma transition in $^8$Be}

\author{V.M. Datar$^1$, Suresh Kumar$^1$, D.R. Chakrabarty$^1$, V. Nanal$^2$,
E.T. Mirgule$^1$, A. Mitra$^1$, and H.H. Oza$^1$}

\affiliation{
$^1$Nuclear Physics Division, Bhabha Atomic Research Centre, Mumbai-400 085,
India\\$^2$Tata Institute of Fundamental Research, Colaba, Mumbai-400 005,
India}
\date{\today}

\begin{abstract}

The  low lying states in $^8$Be are believed to have a two-alpha cluster
structure  and  hence  a  large  intrinsic quadrupole deformation.   An   earlier
calculation  showed  a  large collective enhancement in gamma transition
probability between the low lying states leading to  a  4$^+$  to  2$^+$
gamma  branch of $\sim$10$^{-7}$ and a resonant radiative cross section 
of 134~nb for the $\alpha+\alpha$ entrance channel. We report here the 
first  experimental  evidence  for  this  transition through  a  
$\gamma-\alpha-\alpha$  coincidence  measurement in the  
reaction  $^4$He($\alpha,\alpha \gamma$)$^4$He using a gas target.  The 
measured cross  sections on and off the 4$^+$ resonance are
165~$\pm$~41~(stat)~$\pm$35~(sys)~nb and
39~$\pm$~25~(stat)~$\pm$7~(sys)~nb, respectively.

\end{abstract}

\pacs{21.10.Ky, 23.20.Lv, 24.30.Gd, 27.20.+n}

\maketitle

The  $^8$Be  nucleus  is  believed to be the simplest composite 
$\alpha$-cluster system. The ground state is 92~keV above the $\alpha-\alpha$
threshold  and  is a narrow resonance while the first and second excited
states are much broader resonances~\cite{ajzenberg} as shown in Fig.~1. These were identified  
through  the phase  shift analysis of $\alpha-\alpha$ elastic scattering data and 
have  been  assigned  as  members  of  the   ground   state   rotational
band~\cite{bm2}. This assignment is primarily  based on their energies and spins
calculated using an  $\alpha$-cluster  model~\cite{buck,  rae}.  A crucial  
observable  in  support  of  this assignment is the enhanced E2
transition probability between the successive band members arising  from
the  very large intrinsic deformation predicted by the cluster model. 
To date there does not exist any direct  measurement  of
the  electromagnetic  transition  rate  between  the low lying states of
$^8$Be. Such a measurement, though difficult because of  the  small
gamma  branching  ratio, would also be of interest in the context of the
\begin{figure}
\includegraphics[scale=0.35]{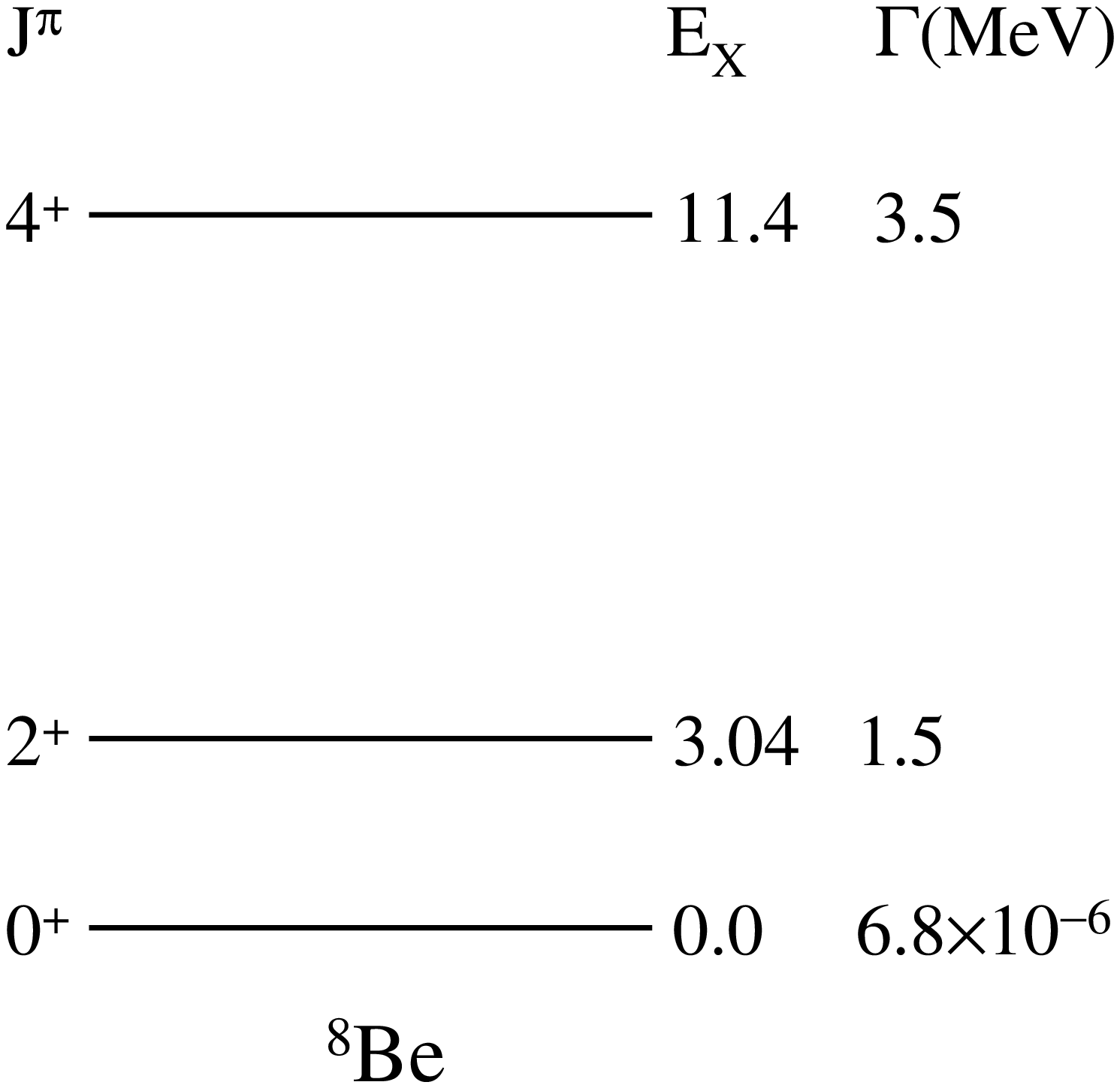}
\caption{Level  scheme  of $^8$Be showing the spin and parity~(J$^\pi$),
excitation energy~(E$_{\rm X}$) and width~($\Gamma$) of
the ground and first two excited states.}
\end{figure}
study of alpha linear chain configurations~(LCC) in  heavier nuclei
such  as  $^{12}$C,  $^{16}$O, $^{20}$Ne and 
$^{24}$Mg~\cite{merchant}. 

There  are  two  measurements in $^8$Be, which is the simplest $\alpha$ LCC, 
searching  for $\alpha-\alpha$  bremsstrahlung  at 
E$_{\alpha}$~=~9.4~MeV~\cite{frois} 
and at E$_{\alpha}$~=~12~-~19~MeV~\cite{peyer}  by
only detecting the charged particles. Coincident $\alpha$~particles were 
measured at fixed angles of $\pm$27.5$^0$ and $\pm$35$^0$,  respectively,  
with  respect  to  the beam.  Frois~{\it  et al.}~\cite{frois} did not find 
evidence for gamma emission and placed an upper limit of 
$\sim$~3~$\mu$b/sr$^2$ at 67\%  confidence level while Peyer~{\it et al.}~\cite{peyer} 
found that the measured cross section is higher than that calculated including 
only  {\it external} bremsstrahlung. The results of Ref.\cite {peyer} suggested 
the presence of an additional {\it internal} bremsstrahlung amplitude 
due  to  the enhanced 4$^+$ to 2$^+$ E2 transition~\cite{green}.  It 
should be mentioned that these measurements were not aimed at measuring, 
even if only indirectly,  the resonant radiative cross section since the beam 
energies were not optimal for either of the 2$^+$ or 4$^+$ $\alpha-\alpha$ 
resonances. A substantial improvement over these measurements 
would be to detect  the  gamma ray  in  coincidence  with  the two alpha particles 
emitted subsequently from the final state. Further, the measurements should be made at  
beam energies corresponding to on and off resonance.

It may be worthwhile at this point to summarize the relevant properties 
of  the  low  lying  states  of $^8$Be.  The  experimental  energies  
and  widths  of  the  lowest 
0$^+$, 2$^+$ and 4$^+$ states can be reproduced reasonably by the 
$\alpha$-cluster model~\cite{buck,  rae}. The widths arise primarily from the $\alpha$-decay
channel since the proton and neutron decay are not energetically possible. 
Using this model the E2 gamma decay widths 
for the first two excited states  of  $^8$Be have been calculated~\cite{langanke1,
langanke2}  to  be  8.3~meV and 0.46~eV corresponding to  the 
B(E2)  values  of $\sim$~75~W.u.  and  $\sim$~19~W.u.~\cite{note}, respectively, 
indicating a substantial collective enhancement. However, the calculated
\begin{figure*}
\includegraphics[scale=0.45]{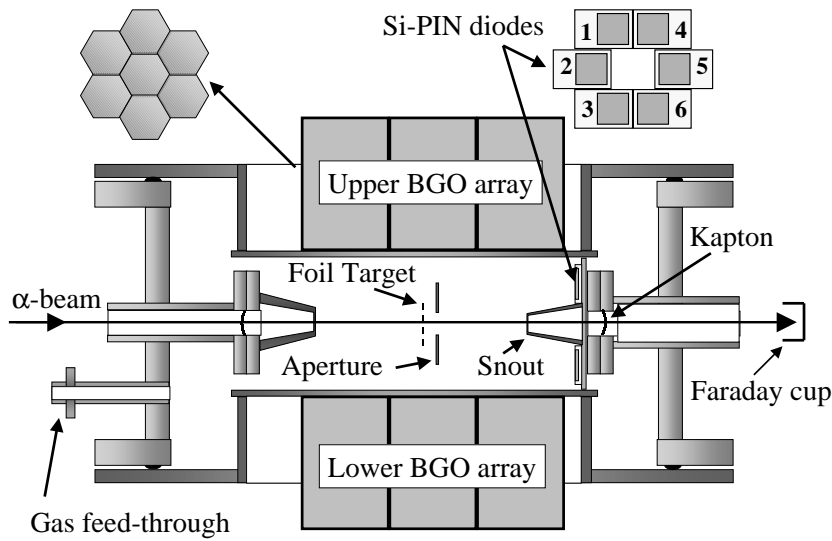}
\caption{A  schematic  of  the  experimental  setup.}
\end{figure*}
gamma branching ratios for the two transitions are still very small 
{\it viz.}~5.9~$\times$~10$^{-9}$ and 1.2~$\times$~10$^{-7}$, respectively.
The resonances in the reaction 
$^4$He($\alpha,\gamma$)$^8$Be are predicted at excitation energies 
of 2.8 MeV and 10.7 MeV, which are lower than the energies shown 
in Fig.~1, with the corresponding peak cross sections of 14~nb  and  134~nb. 
Considering the relatively larger cross section we 
have chosen to measure the gamma decay of the 4$^+$ resonance.
In this letter we present the  first direct experimental evidence for this 
electromagnetic decay branch.

The $^8$Be nucleus was populated through the $\alpha+\alpha$ channel, 
using a gas target, at two beam energies of 22.4 MeV and 26.5 MeV. 
These correspond to on and off the 4$^+$ resonance, respectively, in 
the reaction zone after accounting for the energy loss in the intervening 
material. Gamma rays were measured in coincidence with the 
two $\alpha$~particles from the breakup of the final state. 
The alpha detectors were placed in the forward direction so as to optimise the
detection efficiency for the coincident  $\alpha$~particles which emerge from the
target region within $\sim$30$^\circ$ of the beam. The 4$^+$ to 2$^+$ radiative
transition in $^8$Be has been detected in the present experiment.   
The measured  cross sections are  165~$\pm$~41~(stat)~$\pm$~35~(sys)
and 39~$\pm$~25~(stat)~$\pm$~7~(sys)~nb, for on and off resonance, 
respectively. The measurements agree reasonably with the cluster model 
calculation of Ref.\cite{langanke2}.

The  experiment  was  done   using alpha
beams from the 14UD Pelletron at Mumbai with a beam current of 
$\sim$1~pnA. The reaction chamber was designed for
use of a gas target and housed a charged particle detector array at forward  angle.
It allowed  the  placement  of  gamma  detectors in a compact geometry
leading to a high combined detection efficiency. Bi and C foil targets could be inserted 
in the beam path for calibration of the charged 
particle detectors by measuring the elastic and inelastic scattering of the alpha beam.
A schematic diagram of the set up is  shown  in  Fig.~2.  A  helium  gas
target  of  purity  $>$~99.9\% was used at 0.8~bar pressure and the target 
chamber was isolated from the beam line vacuum  by 
1.8~mg/cm$^2$  
thick  Kapton  foils at the entry and exit. The target 
chamber was evacuated through a liquid nitrogen trap~(LNT) and flushed with
helium passing through another LNT. This cycle  of  evacuation
and  flushing  was repeated a few times before filling the chamber with helium
at the above mentioned pressure.  Gamma  rays
were  detected  in  two arrays, each having seven close packed hexagonal
bismuth    germanate~(BGO)    scintillation detectors
placed  $\sim$~3~cm  above  and below the target. The
upper BGO detectors had a thickness of 76~mm and a face to face  width  of
56~mm  while  the corresponding numbers for the lower set were 64~mm and
56~mm. The  two  $\alpha$~particles  resulting  from  the  decay  of  the  2$^+$
resonance,  subsequent  to  the 4$^+$ to 2$^+$~$\gamma$ transition, were
detected in an array of six silicon~(Si) PIN diodes~(9~mm$\times$9~mm$\times$500~$\mu$m).
These were placed  at  forward  angles in a close packed configuration (see
Fig.~2). A suitably placed 8~mm aperture in the beam path blocked 
elastically scattered alpha particles from  the Kapton  window. The downstream 
snout reduced the coincident elastically scattered  $\alpha$~particles. A Monte Carlo 
simulation  was  used  to optimise the efficiency and decide the aperture 
and Si detector geometry. 
The simulations indicated an effective target length
of about 12~mm centered around the aperture and the angles covered by the
forward Si detectors to be between~$\sim$~15$^{\circ}$
to 35$^{\circ}$. A 300~$\mu$m thick Si detector was placed 
at $\sim$110$^{\circ}$ with respect to the beam and the spectra were recorded 
separately to  monitor the impurities (mainly air) in the helium gas via 
their corresponding elastic peaks. In order to estimate the contribution of
the residual air background, measurements were also 
made  with air at 0.1~bar at both beam energies. This pressure 
results in a  similar  energy  loss for the incoming and outgoing alpha particles 
as with the helium gas target.

The  energy  resolution  of  the  Si  detectors  for  5.5~MeV alpha particles from an 
$^{241}$Am-$^{239}$Pu source was $\sim$~50 keV. The BGO arrays were 
calibrated for high energy gamma rays using $^{241}$Am-Be 
(4.44 MeV) and  $^{238}$Pu-$^{13}$C (6.13 MeV) sources.
The energy resolution for 4.44~MeV gamma 
rays  in  each  array  was $\sim$~7.5$\%$. The gamma ray response of the 
arrays was  simulated using  the  code  GEANT~\cite{cern}, taking into account 
the attenuation due to the stainless steel lids of the target chamber. The 
simulation,  after background  addition, agrees with the measured 
source spectrum in shape and magnitude (within $\pm$~15\%).
\begin{figure}
\includegraphics[scale=0.5]{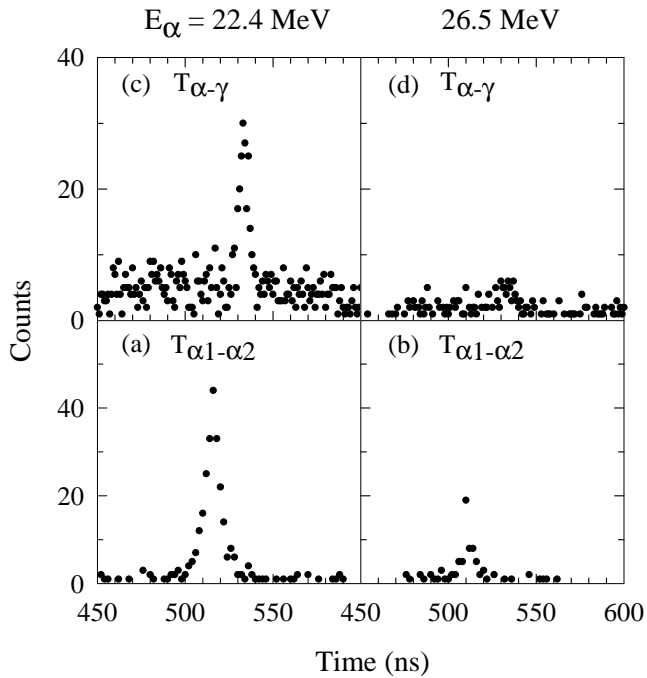}
\caption{Summed time  spectra  between diametrically opposite Si detectors 
(a),~(b) and Si-BGO (c),~(d) gated by 
E$_{\alpha1}$~+~E$_{\alpha2}$~+~E$_{\gamma}$ of 
19$-$22 and 23$-$26 MeV for on and off resonance beam energies, respectively.}
\end{figure}

The  anode  signal from each BGO detector in an array was sent
to a two way passive splitter and the two outputs were sent to  two  separate
fast  amplifiers. One amplifier was set to a higher gain and
 was  used  to  generate  a  fast logic  timing  signal. The output of the other was 
summed and sent to a charge sensitive analog to digital
converter (with a 1~$\mu$sec wide gate)  to  generate the  total energy signal. The signal from 
the charge sensitive preamplifier for each Si detector was processed to give an 
analog energy signal and a  fast timing  signal. Data were collected in the event 
by event  mode, in  a   computer based data   acquisition   system
~\cite{chatterjee}, requiring at least two Si detectors and one of the 
BGO arrays in fast coincidence. The parameters recorded were 1)~the energies 
of the Si detectors, 2)~the sum energies  of  the upper and lower BGO 
arrays and 3)~the timing between the BGO arrays and each Si detector. 
The  on  and off resonance data have been acquired for beam
charges of 112~p$\mu$C and 69~p$\mu$C, respectively. 

The data was sorted by selecting the diagonally opposite (1-6, 2-5 or 3-4, see Fig. 2) 
Si detectors in coincidence with each other and with one of the BGO 
arrays. Typical one dimensional projections of the BGO$-$Si time spectra and Si$-$Si
time   spectra   are   shown  in  Fig.~3. Two  dimensional~(2D) 
E$_{\gamma}-($E$_{Si1}$+E$_{Si2}$) spectra, where E$_{\gamma}$  is 
the energy deposited  in  one of the gamma detector arrays and 1 and 2 refer to any
of the three diagonally opposite pairs, were projected  from  the  data.
The  gates  used  during  the  projection  ensured  a prompt coincidence
between the Si detectors, an energy deposit $\geq$~2~MeV in each Si
detector and a prompt Si$-$BGO coincidence with E$_{BGO} \geq$~2~MeV.
The random contribution was estimated from the 2D spectra created 
by putting appropriate prompt and random gates in the Si1$-$Si2 and Si1$-$BGO time 
parameters.  Finally, E$_{\gamma}$  gates  of 5.0~$-$~12.5~MeV~(on resonance)
 and 7.0~$-$~14.5~MeV~(off resonance) 
were used to generate the  sum~(E$_{Si1}$+E$_{Si2}$+E$_{\gamma}$) specra
from the above 2D spectra. 
The background contribution  due to the residual air was 
measured and found to be  small.  Figure~4 shows the final  sum  spectra 
at the two beam energies after subtracting the impurity contributions and 
random coincidences.  A peak can be seen
in the  E$_{Si1}$+E$_{Si2}$+E$_{\gamma}$  sum  energy  spectrum  at  the
lower~(on  resonance)  beam  energy. This clearly corresponds to the 
4$^+$  to  2$^+$  radiative transition in $^8$Be. 
The peak energy is lower than the beam energy at the target zone because of 
the energy loss of $\alpha$~particles, following the radiative transition, in 
the intervening gas region before they reach the Si detectors.
At the higher~(off resonance) beam energy there is a small excess 
of counts above  background  but  with  a
larger fractional error. Similar sum spectra have also been generated for non 
diagonal pairs (1-5, 2-6 and 3-5) of Si detectors in coincidence 
with the gamma detectors. The on resonance spectra do not show such a 
peak. This is as expected because the two alpha particles and the beam have to lie in the
same plane since the recoil effect due to the radiative transition is negligible. 
This observation also shows that there is no significant contribution from events 
corresponding to a sequential $\alpha$ emission from the excitation of the target 
impurities accompanied by pileup in the BGO array leading to a peak in the 
energy region of interest.
\begin{figure}
\includegraphics[scale=0.53]{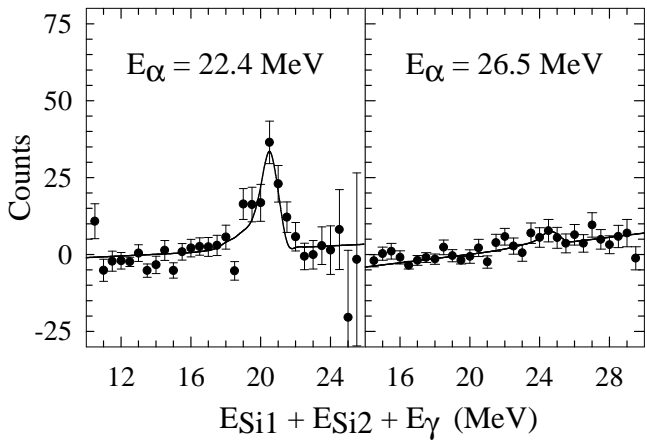}
\caption{Sum energy spectra of E$_{Si1}$+E$_{Si2}$+E$_{\gamma}$ on and 
off the resonance after random and air background subtraction along with the 
simulation for the best fit radiative cross section.}
\end{figure}

In order to extract the cross section for the radiative transition 
in the present reaction, a  Monte  Carlo
program  was  used  to simulate the experiment and generate 
$\gamma-\alpha-\alpha$ events. The   angular
distribution of the alpha particles emitted following E2 gamma decay was included
in  the simulation. 
The measured alpha  singles  spectral shapes
were compared with the Monte Carlo simulations to constrain the possible 
transverse  position  offsets of the beam. These were found to be
$<$~1~mm and result in a change in the triple coincidence efficiency of 
less than $\pm$~5\%.  
The gamma ray response for the two BGO arrays were  simulated  using  the  code
GEANT,  taking  into  account  the angular distribution of the
E2 gamma rays. Typical values of the photopeak efficiency (defined for a
deposited energy of upto 1 MeV lower than  the  full  energy  peak)  are
about  12.5\%  and  11.5\%  for  gamma  ray  energies  of  8 and 10~MeV,
respectively for each of the BGO arrays. This response was used in the simulation 
program to generate a data file from which the 
E$_{Si1}$+E$_{Si2}$+E$_{\gamma}$ spectra were projected. These were scaled 
and added to a linear background to fit the data. After 
accounting for the efficiency loss due to the dead time of the acquisition system and 
the gates used in the data analysis, this scaling factor was used to obtain the best 
fit cross section. The best fits to the data, on and off the resonance, are 
shown in Fig.~4.  The extracted gamma  cross sections at   the two beam energies 
of 22.4 and 26.5~MeV are 165~$\pm$~41~(stat)~$\pm$~35~(sys)  and 
39~$\pm$~25(stat)~$\pm$~7~(sys)~nb, respectively. The systematic error 
includes the error due to the uncertainty in beam position, the choice 
of the fitting procedure and the subtracted air background. The measured on and off 
resonance gamma cross   sections compare favorably with the 134~nb 
and $\sim$12~nb  calculated  in Ref.\cite{langanke2} within experimental 
errors. This corresponds to a B(E2) of  
24~$\pm$~6~(stat)~$\pm$~5~(sys)~e$^2$~fm$^4$ assuming a 
Breit-Wigner shape for the resonance as in Ref.\cite{langanke2}. 
This also agrees, within errors, with the B(E2) of 18.2~$\pm$~0.4 e$^2$~fm$^4$ 
calculated by the {\it  ab initio} quantum Monte Carlo method~\cite{wiringa}. 
 
The  natural extension to the present work would be to scan the 4$^+$
resonance and to search for the 2$^+$  to  0$^+$  gamma
transition  in  $^8$Be. A more precise measurement than the one
described above would require higher segmentation of  the 
alpha  detectors in order to handle higher beam currents. This will 
also give information on the angular correlation of the emitted 
$\alpha$~particles after the radiative transition. 

In  summary  we have made the first direct observation of the 4$^+$ to
2$^+$ gamma transition in $^8$Be through a $\gamma-\alpha-\alpha$
triple  coincidence  measurement.  The  derived on and off resonance 
gamma cross sections agree reasonably with an earlier cluster model calculation.
The present experiment demonstrates the feasibility of gamma ray measurements 
to confirm the $\alpha$~LCC in heavier nuclei. 

We  thank  the Pelletron crew for delivering the alpha beam, P. Sugathan
for providing the Si detectors, S.~Rathi, R.G.~Pillay, A.~Chatterjee 
and P.V.~Bhagwat for their help during the experiment.

\end{document}